\documentclass[a4paper]{article}
\usepackage{hyperref}
\usepackage[dvipsnames]{xcolor}
\usepackage{multirow}
\usepackage{dblfloatfix}

\usepackage{makecell}

\usepackage{INTERSPEECH2019}

\title{WeSinger: Data-augmented Singing Voice Synthesis with Auxiliary Losses}

\name{Zewang Zhang, Yibin Zheng, Xinhui Li, Li Lu}

\address{
  Tencent Inc, Guangzhou, China}

\email{\{zewangzhang, hiccupli\}@tencent.com}

\begin{document}

\maketitle
\begin{abstract}

In this paper, we develop a new multi-singer Chinese neural singing voice synthesis (SVS) system named WeSinger. To improve the accuracy and naturalness of synthesized singing voice, we design several specifical modules and techniques:  1) A deep bi-directional LSTM-based duration model with multi-scale rhythm loss and post-processing step; 2) A Transformer-alike acoustic model with progressive pitch-weighted decoder loss; 3) a 24 kHz pitch-aware LPCNet neural vocoder to produce high-quality singing waveforms; 4) A novel data augmentation method with multi-singer pre-training for stronger robustness and naturalness. To our knowledge, WeSinger is the first SVS system to adopt 24 kHz LPCNet and multi-singer pre-training simultaneously. Both quantitative and qualitative evaluation results demonstrate the effectiveness of WeSinger in terms of accuracy and naturalness, and WeSinger achieves state-of-the-art performance on the recent public Chinese singing corpus Opencpop\footnote{https://wenet.org.cn/opencpop/}. Some synthesized singing samples are available online\footnote{https://zzw922cn.github.io/wesinger/}.

\end{abstract}

\section{Introduction}
Recently, the SVS task has drawn increasing attention for its various potential applications in entertainment and multi-model technologies. Unlike the traditional text-to-speech (TTS) pipeline, a robust SVS (singing voice synthesis)  system aims to generate accurate and natural singing voices from linguistic and musical features such as the phoneme, tempo, pitch, slur, and duration. A typical SVS system consists of three essential components: an acoustic model to convert lyrics and musical information into acoustic features, a duration model to predict the duration of each phoneme, and a neural vocoder model to generate singing voices from acoustic features. In the past, some SVS systems have been designed on the unit selection~\cite{macon1997singing,bonada2016expressive} or statistical parametric speech synthesis (SPSS) methods like context-dependent hidden Markov model (HMM)~\cite{oura2010recent}. Nevertheless, the quality of synthesized singing voices 
cannot reach the naturalness of ground-truth songs owing to the over-smoothing effects and lack of generalization.

\iffalse
\begin{figure}[htp]
    \centering
    \includegraphics[width=8cm]{general_architecture_of_SVS_v3.pdf}
    \caption{An overall pipeline of WeSinger with different components and stages.}
    \label{fig:pipeline}
\end{figure}
\fi

Thanks to the emergence of deep neural network (DNN)~\cite{lecun2015deep} and some significant gains obtained in TTS, including popular acoustic models~\cite{wang2017tacotron,ren2019fastspeech} and waveform generation algorithms~\cite{kalchbrenner2018efficient,valin2019lpcnet,kong2020hifi}, a great deal of improved DNN-based models have also been suggested subsequently for acoustic modeling in SVS systems~\cite{nishimura2016singing,blaauw2017neural,nakamura2019singing,hono2021sinsy}. XiaoiceSing~\cite{lu2020xiaoicesing} employs acoustic model inspired by FastSpeech~\cite{ren2019fastspeech} with a tradition vocoder World~\cite{morise2016world} to outperform the conventional DNN-based systems. HiFiSinger~\cite{chen2020hifisinger}, N-Singer~\cite{lee2021n} and CpopSing~\cite{wang2022opencpop} are equipped with Transformer-like acoustic model and GAN-based neural vocoder to achieve accurate and high-fidelity singing synthesis. In recent literature, another fast GAN-based multi-singer singing voice vocoder Multi-Singer~\cite{huang2021multi} is proposed to improve the performance of singing voice synthesis for the unseen singers. 
Apart from the feed-forward paradigm, an auto-regressive Chinese SVS system ByteSing~\cite{gu2021bytesing} is designed based on the Tacotron-based~\cite{wang2017tacotron} acoustic models with attention mechanism and WaveRNN~\cite{kalchbrenner2018efficient} neural vocoder. Moreover, the pipeline of building data directly mined from the web without any high-quality singing data is also explored in DeepSinger~\cite{ren2020deepsinger}. 

%One prerequisite for a good SVS system is that the rhythm naturalness and pitch accuracy should be reflected in the synthesized singing voices. 
Although previous attempts bring different degrees of performance improvements in acoustic modeling or parallel neural vocoder for SVS tasks, they pay little or no attention to better rhythm modeling, such as the dependency between syllables and phonemes in Chinese. Meanwhile, previous works either adopt Mel-spectrogram without introducing pitch information explicitly as the target for acoustic modeling or consider pitch information as acoustic feature input for GAN-based neural vocoder. As a result, the acoustic model ignores the importance of pitch for SVS and may cause 
the predicted F0 contour out of tune, while the GAN-based neural vocoders may produce some unsatisfactory artifacts in the quality of synthesized singing voices especially when the human ear is sensitive to the coherence of sound. In addition, how to utilize multi-singer pre-training with data augmentation method to improve naturalness and robustness for SVS tasks has also not been explored yet.

\begin{figure*}[htp]
    \centering
    \includegraphics[width=17cm]{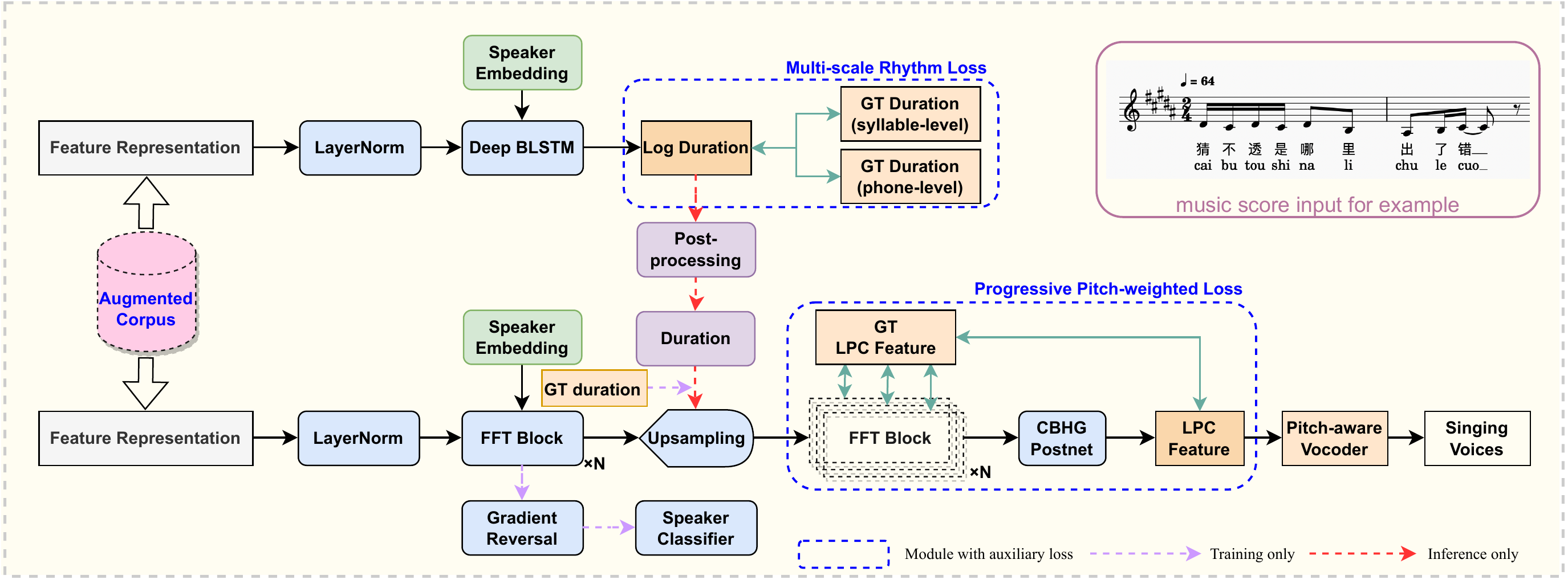}
    \caption{The system architecture of our proposed SVS system WeSinger. First, each song recording is collected with MusicXML-format~\cite{good2001musicxml} transcription and human-annotated phoneme-level interval file. Then, a data augmentation method of variable-length segmentation is applied to attain a larger multi-singer corpus. The training stage contains multi-speaker pre-training and fine-tuning with data of a specific singer. At the synthesis stage, WeSinger first predicts the duration of each phoneme by duration model with a post-processing procedure and then produces pitch-aware LPC features as input for the pitch-aware 24 kHz LPCNet vocoder.}
    \label{fig:wesinger}
\end{figure*}

To address the above challenges and meet the demands of naturalness in rhythm and accuracy in pitch, we propose a new end-to-end SVS system named WeSinger to imitate the voice and style of any singer. Our contributions are summarized below. First, we design two auxiliary loss functions for optimizing the duration model and acoustic model effectively, including multi-scale rhythm loss and progressive pitch-weighted loss. Second, to achieve the high-quality and pitch-aware synthesis, we adopt a 24 kHz pitch-aware LPCNet vocoder to convert acoustic features into singing waveforms. Last, we propose a data augmentation method and combine the augmented multi-singer corpus with pre-training and then do fine-tuning to further enhance the performance of WeSinger. Experimental results indicate that the data-augmented WeSinger with auxiliary losses can synthesize singing voices with much higher quality.

\section{Methodology}
\label{section2}

\subsection{Architecture}
The overall architecture of WeSinger is described in Figure~\ref{fig:wesinger}, each module of which will be introduced in detail.

\subsubsection{Feature Representation}
Given MusicXML-format~\cite{good2001musicxml} files and human-annotated phoneme-level interval files, we first convert the Chinese syllable sequences in lyrics to phoneme sequences using the Chinese grapheme-to-phoneme algorithm and map the combination of \emph{step}, \emph{octave} and \emph{alter} into a single note pitch according to the MIDI standard~\cite{midi1996complete}. Chinese syllables are usually composed of initials and finals. In this case, we build phoneme-type sequences from phoneme sequences for richer feature representation. The duration of each syllable is calculated by its associated notes and beats per minute (BPM), and we observed that it works best to divide the duration evenly between the syllable's initial and final. Notably, we don't make use of 
Chinese tone information due to the existence of musical notes. Furthermore, we assign each phoneme a slur type from \emph{start}, \emph{continue}, \emph{stop} and \emph{null} to indicate whether to be played smoothly or not. The ground-truth duration of each phoneme is extracted from the annotated interval file.
All durations are converted to the number of frames with a unit of 10 milliseconds. Finally, we rearrange all of the pitch sequence, slur sequence, beat sequence, phoneme-type sequence, and duration sequence to manually match the phoneme sequence. For convenience, we abbreviate some input symbols and illustrate them in Table~\ref{tab:symbol_description}.

\begin{table}[]
\centering
\caption{Abbreviations and descriptions of some input symbols in WeSinger.}
\begin{tabular}{l l}
\hline
Abbreviation         & Description    \\ \hline \hline
\emph{\textbf{Ph}}          & Chinese phoneme (ch, ang) \\
\emph{\textbf{Pt}}          & Chinese phoneme type (initial, final,\\  & single final) \\ \hline
\emph{\textbf{Pi}}          & Note pitch (C4, D4) \\
\emph{\textbf{Sr}}          & Slur flag (start, continue, stop, null) \\ \hline
\emph{\textbf{Bt}}          & Duration of note beat (frames) \\ \hline
\end{tabular}
\label{tab:symbol_description}
\end{table}

\subsubsection{Duration Model}
 One discrepancy between TTS and SVS is that the starting and ending time of each phoneme in SVS systems is largely constrained by music scores, while traditional TTS systems often do duration prediction with higher degrees of freedom. We evaluate the different forms of feature representation and finally take the most informative and effective phoneme-level $ \textbf{X}_D = [\emph{\textbf{Ph}}, \emph{\textbf{Pt}}, \emph{\textbf{Pi}}, \emph{\textbf{Sr}}, \emph{\textbf{Bt}} ]$ as the input of BLSTM-based duration model and log-scale duration as the target. We also find that it can make training more stably and efficiently by adopting the layer normalization~\cite{ba2016layer} technique for the concatenated input embedding sequences. To make the rhythm of synthesized singing voice more accurate and natural,
 we design a multi-scale rhythm loss including the commonly used phoneme-level minimum mean squared error (MMSE) loss and an auxiliary syllable-level MMSE loss to optimize the duration model. Normally, for a syllable sequence ($s_1$, $s_2$, ..., $s_M$) and its phoneme-level sequence ($p_1$, $p_2$, ..., $p_N$), if we denote the ground-truth phoneme-level duration as $d_{p_i}$ and the predicted phoneme-level duration as $\hat{d}_{p_i}$, thus, the loss function of the duration model can be written as:
 \begin{equation}
 \mathcal{L}_{dur} = 
 \frac{1}{M}\sum_{i=1}^{M} {\lvert \sum_{p_k \in s_i} (d_{p_k} - \hat{d}_{p_k}) \lvert}^2 
 +\frac{1}{N}\sum_{j=1}^{N} {\lvert d_{p_j} - \hat{d}_{p_j} \lvert}^2 \\
\end{equation}

Indeed, there are time-lags~\cite{blaauw2017neural} between the starting time of musical notes and the starting time of ground-truth audios. To enforce the constraint that the duration of a syllable should be consistent with musical notes, we adopt a post-processing procedure to refine the predicted duration by scaling the predicted duration of phonemes in each syllable in proportion to the duration of corresponding musical notes, limiting the duration of its consonant to a maximum of 100 milliseconds according to the duration distribution of the initials in the training set, and allocating the remaining duration to the final.

\subsubsection{Acoustic Model}
The acoustic model of WeSinger consists of an encoder-decoder module modified from FastSpeech~\cite{ren2019fastspeech}. A carefully designed phoneme-level $ \textbf{X}_D = [\emph{\textbf{Ph}}, \emph{\textbf{Pt}}, \emph{\textbf{Pi}}, \emph{\textbf{Sr}} ]$ is represented as the input for encoder. Apart from adopting the layer normalization technique for the concatenated embedded input sequences and substituting the intermediate fully-connected layers of each Transformer FFT block with Conv1D layers for stronger context-dependent performance\cite{gulati2020conformer}, we made two key improvements for better acoustic modeling. First, to discourage the encoder from also memorizing speaker characteristics in multi-singer pre-training and improve the generalization, we employ the domain adversarial training (DAT) strategy~\cite{ganin2016domain} by appending a speaker classifier with a gradient reversal layer (GRL)~\cite{zhang2019learning, Zheng2022Zero} to the encoder outputs. We find it effective to set the weight for GRL loss as 0.02. To our knowledge, this is the first time that GRL is successfully applied to the training of a multi-singer SVS system. Second, in contrast to the commonly-used Mel-spectrogram as the target~\cite{ chen2020hifisinger,gu2021bytesing}, we choose to adopt the pitch-aware features including Bark-scale Frequency Cepstral Coefficients (BFCCs) and pitch information as the intermediate target for WeSinger. We design a progressive pitch-weighted loss function for optimizing WeSinger's acoustic model. Specifically, for output states of all the FFT blocks and post-net in the decoder, we insert a fully-connected layer to project the intermediate states to the target feature space and calculate the L1 loss between the projected feature and the target acoustic feature. Besides, we observe that re-weighting the L1 loss by multiplying 1.2 in the dimension of log-scale F0 can achieve better pitch prediction without harming pronunciation. Normally, given the sequence length of the decoder's output as $T$, the number of all decoder blocks as $B$, the target LPC feature sequence as $Y$, the 26-dim weighting vector as  $w=[1,1,...,1.2,1]$, and the transformation functions as $\mathcal{F}$, then the formula for the decoder's loss function can be written as follows:
\begin{equation}
%\[ 
 \mathcal{L}_{decoder} = 
 \frac{1}{T}\sum_{i=1}^{T} \frac{1}{B}  \sum_{j=1}^{B} 
 w \circ {\lvert \mathcal{F}_j(state_{i,j}) - {Y}_{i} \lvert} 
 %\]
 \end{equation}

\iffalse
\begin{figure}[htp]
    \centering
    \includegraphics[width=8cm]{data_augmentation_SVS.pdf}
    \caption{Illustration of PS strategy.}
    \label{fig:pitchshift}
\end{figure}
\fi

\subsubsection{Neural Vocoder}
Despite the recent popularity of parallel GAN-based vocoders~\cite{kong2020hifi}, we still find some unsatisfactory artifacts in the quality of synthesized singing voices~\cite{chen2020hifisinger,lee2021n,huang2021multi} especially when the human ear is sensitive to the coherence of sound. Different from ByteSinger~\cite{gu2021bytesing} which adopts WaveRNN~\cite{kalchbrenner2018efficient} as a neural vocoder without modeling the pitch information explicitly, we design a pitch-aware 
neural vocoder to synthesize singing waveforms with a sampling rate of 24 kHz. Our pitch-aware neural vocoder is modified from the LPCNet~\cite{valin2019lpcnet}, which was originally proposed as a neural vocoder for TTS tasks to synthesize 16 kHz speech in an auto-regressive manner. Considering the widely held view that a lower sampling rate as 16 kHz is far away from representing singing voices more accurately, we expand the 18-dimensional BFCCs in the original LPCNet to 24-dimensional BFCCs and combine 24-dimensional BFCCs with two pitch parameters (log-scale F0 and pitch correlation)
as input features for our proposed 24 kHz pitch-aware neural vocoder. In the meantime, we increase the dimension in the hidden state of GRU$_A$ from the original 384 to 512 and keep that of GRU$_B$ unchanged. To the best of our knowledge, this is the first attempt to adopt an auto-regressive pitch-aware neural vocoder for the Chinese SVS tasks.

\subsection{Data Augmentation}
\label{da}
It is challenging to obtain an exhaustive singing dataset that sufficiently covers the diversity of rhythms, lyrics, and melodies. Furthermore, it is a costly and laborious task to gather correct annotations. In WeSinger, we try to alleviate such a dilemma by an effective data augmentation method of variable-duration segmentation (\emph{VS}). 
 Since the acoustic model of WeSinger is based on the Transformer architecture with MHSA having a global receptive field, in which the modeling ability is inevitably sensitive to the length of the input sequence. Under this perspective, we segment each song recording into smaller fragments of three different time intervals: $0 \sim 5$ seconds, $5 \sim 8$  seconds, and $8 \sim 12$ seconds. That is to say, each song recording would be split into short audio clips three times. By the way, we also tried the other data augmentation method such as pitch shifting. Specifically, the overall pitch of each song is raised by one semitone and lowered by one semitone in turn. A similar approach has been proposed in~\cite{blaauw2017neural}. This allows us to obtain two additional distinct variants based on each training audio clip. However, we find that doing so could lead to somewhat perceptible changes in timbre. Based on the existing \emph{VS} method,  we found no significant benefit from adopting the additional pitch-shifting method. Therefore, in this paper, we only present the performance of adopting the \emph{VS} method.

\section{Experiments}
\subsection{Experimental Setup}
\subsubsection{Dataset}

We collect three datasets for SVS experiments totally: 
\begin{enumerate}
\item a 30-hour singing dataset collected from nearly 160 amateur singers in different noisy environments.
\item Opencpop~\cite{wang2022opencpop}, a recent public Mandarin singing corpus including a pre-defined training set and test set.
\item a 5-hour internal high-quality Mandarin singing dataset collected from a professional female singer. We pick 5 songs as the test set for our experiments and it does not overlap with all other training sets.
\end{enumerate}
All songs are down-sampled to 24 kHz with 16-bit quantization. All 26-dim acoustic features are re-scaled by min-max normalization for acoustic modeling and then de-normalized for training the LPCNet vocoder. Normally, each song is split into singing segments by the pause, these segments last from 1 second to 5 seconds (4 seconds on average). As for the \emph{VS}-augmented data, each song is split into different intervals as described in Section~\ref{da}.

\subsubsection{Experiments}
Our experiments are conducted in two groups. One group is to explore the optimal performance following different training recipes with only the data of the professional female singer. Based on the conclusion drawn from the first group, we further perform the other group to enhance performance with the help of multi-singer pre-training and data augmentation. In detail, we design several experiments under different conditions as below: 
\begin{itemize}
  \item \textbf{WeSinger} Training the proposed WeSinger system with multi-scale duration loss and progressive pitch-weighted decoder loss.
  \item \textbf{WS-$w/o$-syllable} Training WeSinger without auxiliary syllable loss for duration model.
  \item \textbf{WS-$w/o$-weighted} Training WeSinger without weighted loss on pitch dimension.
  \item \textbf{WS-$w/o$-progressive} Training WeSinger without progressive loss.
  \item \textbf{WS-$w/$-\emph{VS}} Training WeSinger with additional \emph{VS}-augmented data.
  \item \textbf{WS-Finetune} Pre-training WeSinger on multi-singer data and then do fine-tuning.
  \item \textbf{WS-$w/$-\emph{VS}-Finetune} Pre-training WeSinger on multi-singer data with additional \emph{VS}-augmented data and then do fine-tuning.
\end{itemize}
To compare the performance of WeSinger with CpopSing in ~\cite{wang2022opencpop}, we further conduct two experiments following the above training recipes of WeSinger and WS-$w/$-\emph{VS}-Finetune based on the public corpus Opencpop. 

\begin{table}[]
\centering
\caption{Duration Accuracy of with and without auxiliary syllable loss.}
%\resizebox{\linewidth}{!}{
\begin{tabular}{ c c c c c}
\hline
System & \makecell{Dur Acc} &  \makecell{Dur CORR}    \\ \hline
WS-$w/o$-syllable& $87\%$ & $0.96$\\ 

WeSinger & $88\%$ & $0.97$ \\ \hline
\end{tabular} 
%}
\label{tab:acc_duration}
\end{table}

\begin{table}[]
\centering
\caption{Quantitative performance of different SVS systems on an internal professional female singer corpus.}
\resizebox{\linewidth}{!}{
\begin{tabular}{ c c c c c}
\hline
System & \makecell{F0 \\ RMSE} & \makecell{F0 \\ CORR} &\makecell{V/UV \\ error} & \makecell{BFCCD}    \\ \hline

WS-$w/o$-weighted&14.5 & 0.98 & 0.06 & 64.2   \\ 
WS-$w/o$-progressive&15.3 &0.97 &0.06 &66.7 \\ \hline
WeSinger & 13.9 & 0.98 & 0.06 & 63.8  \\ 
WS-$w/$-\emph{VS}&14.1 &0.98 &0.05 &63.5 \\ \hline
WS-Finetune&13.6 &0.98 &0.05 &63.3 \\
WS-$w/$-\emph{VS}-Finetune&12.9 &0.98 &0.05 &63.7 \\ 
\hline
\end{tabular}
}
\label{tab:evaluation_results}
\end{table}

\subsection{Evaluation}
\label{evaluation}
In both quantitative and qualitative tests, we keep the rhythm and lyrics of the test set consistent among different models to examine the audio quality. Notably, the generated short singing voices are combined into each submitted singing segment for evaluation.

%Apart from the architecture designs including multi-scale duration loss and progressive pitch-weighted loss, we described two DA strategies for multi-singer training in section~\ref{da}, so we also want to investigate how different fusions of DA strategies improve the audio quality of SVS. 

\subsubsection{Quantitative evaluation}
We conduct several quantitative tests to compare different systems. First, we try to verify the effectiveness of using multi-scale duration loss. As shown in Table~\ref{tab:acc_duration}, when the auxiliary syllable duration loss is adopted, the duration accuracy has been improved from 87\% to 88\%. Then, to measure the quality of synthesized singing voices, we adopt F0 RMSE (F0 root mean square error), F0 CORR (F0 correlation), and V/UV error (voice/unvoiced error rate), and BFCCD (BFCC distortion) as objective metrics to calculate. For brevity, the ground-truth durations are used as the target duration for systems to synthesize. The calculated results are listed in Table~\ref{tab:evaluation_results} and we can 
come to the following conclusions: 1) With the progressive pitch-weighted loss, the F0 RMSE can be reduced from 15.3 to 13.9; 2) With \emph{VS}-augmented data, the quality of predicted BFCC can be slightly improved; 3) Combining \emph{VS}-augmented data with multi-singer pre-training, the optimal F0 RMSE of 12.9 can be obtained. Overall, data-augmented WeSinger with auxiliary losses can achieve better performance in quantitative metrics.

\subsubsection{Qualitative evaluation}
 To compare the performances of different systems qualitatively, we conduct the Mean Opinion Score (MOS) evaluations for naturalness and audio quality on the test set. We collected synthesized samples from different systems based on both ground-truth duration and predicted duration. Twenty listeners were asked to rate the quality of each singing voice segment on a scale from 1 to 5, in which 1 means very bad and 5 means excellent. As shown in Table~\ref{tab:mos}, when training from scratch, a significant gain of 0.3 in MOS is obtained with the \emph{VS}-augmented data. With multi-singer pre-training, we also benefit from the \emph{VS}-augmented data and achieve the 3.48 MOS finally. As for the public Opencpop corpus, WeSinger also outperforms CpopSing with a MOS margin of 0.1, and WS-$w/$-\emph{VS}-Finetune achieves a final 3.60 MOS. The detailed evaluation result between WeSinger and CpopSing is listed in Table~\ref{tab:opencpop}, which indicates that WeSinger has achieved state-of-the-art performance on the public corpus Opencpop.

\begin{table}[]
\centering
\caption{MOS ratings for the internal professional female singer corpus with the confidence interval 95\%.}
%\resizebox{\linewidth}{!}{
\begin{tabular}{ c c c c c}
\hline
\multirow{2}{*}{System} & \multicolumn{2}{>{\centering\arraybackslash}m{3.65cm}}{MOS} \\ \cline{2-3}

& \makecell{predicted \\ duration}&  \makecell{ ground-truth\\ duration}     \\ \hline

WeSinger & $2.91\pm0.11$ & $3.02\pm0.08$  \\ 
WS-$w/$-\emph{VS}&$3.18\pm0.08$ & $3.42\pm0.08$  \\ \hline
WS-$w/$-Finetune&$3.40\pm0.09$ &$3.46\pm0.06$ \\
WS-$w/$-\emph{VS}-Finetune&$3.48\pm0.06$ &$3.62\pm0.05$ \\ 
%WS-$w/$-\emph{PS}-FT&3.43 &3.45  \\
%WS-$w/$-\emph{VS}+\emph{PS}-FT&3.45 &3.51  \\ 
\hline
Ground Truth & \multicolumn{2}{>{\centering\arraybackslash}m{3.65cm}}{$4.12\pm0.04$} \\ \hline
\end{tabular}
%}
\label{tab:mos}
\end{table}

\begin{table}[]
\centering
\caption{MOS ratings for the public corpus Opencpop with the confidence interval 95\%.}
%\resizebox{\linewidth}{!}{
\begin{tabular}{ c c c c c}
\hline
\multirow{2}{*}{System} & \multicolumn{2}{>{\centering\arraybackslash}m{3.65cm}}{MOS} \\ \cline{2-3}

& \makecell{predicted \\ duration}&  \makecell{ ground-truth\\ duration}     \\ \hline

CpopSing&$3.20\pm0.10$ &$3.35\pm0.15$  \\ \hline
WeSinger & $3.25\pm0.08$ & $3.45\pm0.12$  \\ 

WS-$w/$-\emph{VS}-Finetune&$3.35\pm0.06$ &$3.60\pm0.10$ \\ \hline

Ground Truth & \multicolumn{2}{>{\centering\arraybackslash}m{3.65cm}}{$4.01\pm0.15$} \\ \hline
\end{tabular}
%}
\label{tab:opencpop}
\end{table}

\section{Conclusion}
In this work, we introduced an end-to-end SVS system named WeSinger to bridge the gap between accuracy and naturalness. We described the specifically designed auxiliary loss functions including multi-scale rhythm loss and progressive pitch-weighted decoder loss. Besides, we adopted a data augmentation method of variable-duration segmentation to improve the performance. Experiment results show that the data-augmented WeSinger system can synthesize singing voices with higher quality and naturalness and achieve state-of-the-art performance on the public corpus Opencpop. For future work, we will investigate how to improve the robustness 
efficiently and explore how to adapt to any unseen singer with lower cost.

\vfill\pagebreak

\clearpage %强制分页
\bibliographystyle{IEEEtran}

\bibliography{mybib}

\end{document}